\documentstyle[preprint,eqsecnum,aps]{revtex}

\begin{document}
\draft
\preprint{nucl-th/9901abc}
\title{
Microscopic Calculations of Weak Interaction Rates of Nuclei in Stellar Environment for A = 18 to 100
}
\author{Jameel-Un-Nabi\\ and \\Hans Volker Klapdor-Kleingrothaus}
\address{Max-Planck-Institut f\"ur Kernphysik, 69029 Heidelberg, Germany
}
\date{\today}
\maketitle
\begin{abstract}
We report here the microscopic calculation of weak interaction rates in stellar matter for 709 nuclei with A = 18 to 100 using a generalized form of proton-neutron quasiparticle RPA model with separable Gamow-Teller forces. This is the first ever extensive microscopic calculation of weak rates calculated over a wide temperature-density grid which includes 10$^{7}$ $\leq$ T(K) $\leq$ 30 $\times$ 10$^{9}$ and 10$\leq$ $\rho Y_{e}$ (gcm$^{-3}$) $\leq$ 10$^{11}$, and over a larger mass range. Particle emission processes from excited states, previously ignored, are taken into account, and are found to significantly affect some $\beta$ decay rates. The calculated capture and decay rates take into consideration the latest experimental energy levels and $ft$ value compilations. Our calculation of electron capture and $\beta$-decay rates, in the $fp$-shell, show considerable differences with a recently reported shell model diagonalization approach calculation.
\end{abstract}

\narrowtext
\label{sec:level1}
The weak interaction have several crucial effects in the course of development of a star. They initiate the gravitational collapse of the core of a massive star triggering a supernova explosion, play a key role in neutronisation of the core material via electron capture by free protons and by nuclei and affect the formation of heavy elements above iron via the r-process at the final stage of the supernova explosion (including the so-called cosmochronometers which provide information about the age of the Galaxy and of the universe). The weak interaction also largely determines the mass of the core, and thus the strength and fate of the shock wave formed by the supernova explosion (see, eg., \cite{Kla83,Gro90}). 

Precise knowledge of the terrestrial $\beta$ decay of neutron-rich nuclei is crucial to an understanding of the r-process. Most of these nuclei cannot be produced in terrestrial laboratories and one has to rely on theoretical extrapolations in respect of beta decay properties. The microscopic calculations of weak interaction rates, performed at that time \cite{Sta90a,Hir93} led to a better understanding of the r-process \cite{Kla83}.

The weak interaction rates in domains of high temperature and density scales are of decisive importance in studies of the stellar evolution. A particularly important input which determines both the final electron (or lepton) fraction of the ``iron''-core prior to collapse (i.e., at the presupernova stage) as well as its initial entropy , is the nuclear beta decay and electron capture rates. These reactions not only lead to a change in the neutron-to-proton ratio in the stellar core material but because of the removal of energy by neutrinos produced in the reactions, they cool the core to a lower entropy state. It is therefore important to follow the evolution of the stellar core during its late stages of hydrostatic nuclear burning with a sufficiently detailed nuclear reaction network that includes these weak-interaction mediated reactions.

The first extensive effort to tabulate the nuclear weak interaction rates at high temperatures and densities, where decays from excited states of the parent nuclei become relevant, was done by Fuller, Fowler, and Newmann (FFN) \cite{Ful80} (such rates are referred to as stellar rates throughout this paper). FFN calculated the stellar weak interaction rates over a wide range of densities and temperatures ($10 \leq \rho Y_{e}$ (g cm$^{-3}$) $\leq 10^{11}$ and  $10^{7} \leq$ T(K) $\leq 10^{11}$) for 226 nuclei with masses between A = 21 and 60. The Gamow-Teller (GT) strength and excitation energies were calculated using a zero-order shell model. They also incorporated the experimental data available at that time. For unmeasured transitions, FFN assumed an average log $ft$ value of 5.0.

The FFN rates were then updated, taking into account some quenching of the GT strength by an overall factor of two \cite{Auf94}. These studies were based on the same strategy and formalism as already employed by FFN. Furthermore these authors simulated the low-lying transitions by the same $ft$-value, while FFN adopted specific values for individual nuclei. Later results from \cite{Auf96} implied the need for a more reliable calculation of stellar rates.

Oda et al. (OHMTS) \cite{Oda94} did an extensive calculation of stellar weak interaction rates of $sd$-shell nuclei in the full ($sd$)$^{n}$-shell model space. They also compared their calculated rates with those of FFN and in certain cases they reported differences in the rates up to two orders of magnitude and more. OHMTS calculated weak process rates for 79 nuclei covering isotopes from the $sd$-shell (A = 17 to 39) for both $\beta^{-}$ and $\beta^{+}$ decay directions.

The proton-neutron quasiparticle random-phase-approximation (pn-QRPA) theory \cite{Hal67,Kru84,Mut89} has been shown to be a good microscopic theory for the  calculation of beta decay half-lives far from stability \cite{Mut89,Sta89}. The pn-QRPA was first developed by \cite{Hal67}. Some extension of the model to deformed nuclei was discussed by \cite{Kru84}, while general formulae for the calculation of odd-odd parent nuclei can be found in \cite{Mut89}. Calculations of beta decay rates for all nuclei, at low temperatures, far from stability by microscopic nuclear theory were first performed by \cite{Kla84}, and then complemented and refined by \cite{Ben88,Sta89,Sta90a,Hir93}. Recent studies by \cite{Hom96} have shown that the best extrapolations to cold neutron-rich nuclei far from stability to date still are given by \cite{Sta90a}. The pn-QRPA theory was also successfully employed in the calculation of $\beta^{+}$/EC half-lives of cold nuclei and again good agreement with experimental half-lives was found \cite{Hir93}. The pn-QRPA theory was then extended to treat transitions from nuclear excited states \cite{Mut92}. Keeping in view the success of pn-QRPA theory in calculating terrestrial decay rates, in the present work this extended model was used to calculate for the first time the weak interaction rates in stellar matter using pn-QRPA theory. One of the main advantages of using this formalism is that one can handle large configuration spaces, by far larger than possible in any shell model calculations, and hence can include parent excitation energies over large ranges of 10's of MeV.

In this present work, we considered a model space up to 7 major shells. Particle emission processes from excited states, which were not considered in previous compilations, are taken into account in this work. We specifically calculate 12 different stellar rates for each parent nucleus. These include $e^{\pm}$-capture rates, $\beta^{\pm}$-rates, (anti)neutrino energy loss rates, probabilities of beta-delayed proton (neutron) emission and energy rates of beta-delayed protons (neutrons).  Our calculation of stellar rates for $sd$-shell \cite{Nab99a} nuclei shows significant differences, especially for decay rates, compared to the earlier works of FFN and OHMTS.

During the course of this work, a handful of electron capture and $\beta$ decay rates  were calculated using the shell model diagonalization approach (SMDA) \cite{Lan98,Mar98}.
 However, due to the very large m-scheme dimensions involved, the GT strength distributions were calculated in truncated model spaces (only a model space of 1 major shell was considered). These authors restricted themselves to parent excited states of a few MeV for the calculation of electron capture rates. For the calculation of $\beta$-decay rates they considered parent excited states usually up to 1~MeV and in addition, back resonances (the GT back resonance are states reached by the strong GT transitions in the electron capture process built on ground and excited states, see \cite{Ful80,Auf94}) built on daughter states below 1~MeV.

We, in general, considered a few 100's of initial and final states in our rate calculation. We consider parent excitation energies up to the particle decay threshold, i.e., minimum of $S_{p}$ and $S_{n}$ (after accounting for the effective Coulomb barrier which prevents a proton from being promptly emitted and the uncertainty in calculation of energy levels). This has the effect that our calculated electron capture rates are, in general, suppressed in comparison to the corresponding rates of FFN at high temperatures and densities. The effect is more pronounced for the case of decay rates. A detailed comparison can be found in \cite{Nab99a,Nab99b,Nab99t}. Our results for capture rates are enhanced for odd-A and odd-odd nuclei in comparison to the corresponding SMDA calculation. For even-even nuclei they are suppressed at high temperatures (T9 $>$ 3, where T9 is the temperature in units of 10$^{9}$ K). For the case of decay rates our calculation is, in general, enhanced for the case of odd-A nuclei. In all other cases our calculated rates are suppressed. The degree of suppression and/or enhancement varies with temperature and density. For more quantitative conclusions we refer to \cite{Nab99b}. Our results do not support the claim by \cite{Lan98} that for capture on odd-odd nuclei FFN placed the GT centroid at too low excitation energy. For odd-odd nuclei no experimental information is available for the GT strength distribution, and \cite{Lan98} also did not present a comparison of the corresponding terrestrial rates of odd-odd nuclei with measured half-lives to show the reliability of their calculation. Our calculation is, in general, in good agreement with the FFN calculation for odd-odd nuclei. However, for certain nuclei, FFN rates exceed ours at high densities. Table~I and Table~II compares some of our calculated electron capture and decay rates with earlier calculations. For the sake of reliability of our calculation, a comparison of all calculated terrestrial rates using the pn-QRPA theory, used in the present work, with measured half-lives, wherever possible, have been made and discussed in \cite{Sta90a,Hir93}.  

 The calculated weak interaction rates for 709 nuclei (A = 18 to 100), including also the neutron-rich nuclei which play a key role in the evolution of the stellar core, can be obtained as files on a magnetic tape from the authors on request. For details of the formalism and the calculations we refer to \cite{Nab99a}.

 Some examples of astrophysical application comprising of the new theoretical data set presented here have been discussed in \cite{Nab99c,Nab99d}.

\begin{table}
\caption{Comparison of the calculated electron capture rates with previous works. This Table is similar to Table~I of [19]. $\rho_{7}$ is the density (in units of 10$^{7}$ g cm$^{-3}$), $T_{9}$ is the temperature (in units of 10$^{9}$ K), QRPA denotes our calculated rates, while SM, FFN and Ref [6] denote rates calculated by [19], [5] and [6], respectively. Exponents are given in parenthesis. All rates are given in units of $s^{-1}$.
\label{table3}}
\begin{tabular}{ccccccc}
Nucleus & $\rho_{7}$ & $T_{9}$ & QRPA & SM & FFN & Ref. [6]\\ \hline
$^{56}$Ni & 4.32 & 3.26 & 9.9 (-3) & 1.3 (-2) & 7.4 (-3) & 8.6 (-3)\\
$^{54}$Fe & 5.86 & 3.40 & 1.3 (-5) & 4.2 (-5) & 2.9 (-4) & 3.1 (-4)\\
$^{58}$Ni & 5.86 & 3.40 & 3.7 (-4) & 8.1 (-5) & 3.7 (-4) & 6.3 (-4)\\
$^{56}$Fe & 10.7 & 3.65 & 1.1 (-6) & 2.1 (-6) & 1.0 (-5) & 4.7 (-7)\\ \hline
$^{55}$Co & 4.32 & 3.26 & 8.0 (-2) & 1.6 (-3) & 8.4 (-2) & 5.1 (-2)\\
$^{57}$Co & 5.86 & 3.40 & 1.6 (-3) & 1.3 (-4) & 1.9 (-3) & 3.4 (-3)\\
$^{55}$Fe & 5.86 & 3.40 & 4.8 (-3) & 1.9 (-4) & 5.8 (-3) & 3.8 (-3)\\
$^{59}$Ni & 5.86 & 3.40 & 4.1 (-3) & 4.7 (-4) & 4.4 (-3) & 4.4 (-3)\\ 
$^{59}$Co & 10.7 & 3.65 & 4.9 (-4) & 7.8 (-6) & 2.1 (-4) & 2.1 (-4)\\
$^{53}$Mn & 10.7 & 3.65 & 1.4 (-2) & 3.3 (-4) & 3.8 (-3) & 5.6 (-3)\\ \hline
$^{56}$Co & 5.86 & 3.40 & 3.3 (-2) & 1.7 (-3) & 6.9 (-2) & 5.1 (-2)\\
$^{54}$Mn & 10.7 & 3.65 & 7.5 (-4) & 3.1 (-4) & 4.5 (-3) & 1.1 (-2)\\
$^{58}$Co & 10.7 & 3.65 & 3.4 (-3) & 3.5 (-4) & 9.1 (-3) & 2.1 (-2)\\ 
$^{56}$Mn & 33.0 & 4.24 & 1.1 (-2) & 1.0 (-4) & 4.1 (-4) & 2.0 (-3)\\
$^{60}$Co & 33.0 & 4.24 & 2.0 (-3) & 1.7 (-4) & 1.1 (-1) & 6.1 (-2)\\ 
\end{tabular}
\end{table}
\begin{table}
\caption{Comparison of the calculated beta decay rates with previous works. This Table is similar to Table~II of [19]. $\rho_{7}$ is the density (in units of 10$^{7}$ g cm$^{-3}$), $T_{9}$ is the temperature (in units of 10$^{9}$ K), QRPA denotes our calculated rates, while SM, FFN and Ref [6] denote rates calculated by [19], [5] and [6], respectively. Exponents are given in parenthesis. All rates are given in units of $s^{-1}$. FFN did not calculate rates for nuclei with A $>$ 60.
\label{table4}}
\begin{tabular}{ccccccc}
Nucleus & $\rho_{7}$ & $T_{9}$ & QRPA & SM & FFN & Ref. [6]\\ \hline
$^{56}$Fe & 5.86 & 3.40 & 5.5 (-13) & 3.9 (-11) & 2.3 (-10) & 5.9 (-11)\\
$^{54}$Cr & 5.86 & 3.40 & 1.8 (-8) & 2.2 (-7) & 2.2 (-5) & 1.5 (-7)\\
$^{58}$Fe & 10.7 & 3.65 & 7.3 (-9) & 5.2 (-8) & 2.6 (-6) & 1.5 (-7)\\
$^{60}$Fe & 33.0 & 4.24 & 1.1 (-5) & 1.7 (-4) & 4.6 (-3) & 1.0 (-3)\\ 
$^{52}$Ti & 33.0 & 4.24 & 1.8 (-5) & 1.3 (-3) & 1.1 (-2) & 1.2 (-4)\\ \hline
$^{59}$Fe & 33.0 & 4.24 & 6.2 (-5) & 6.0 (-5) & 6.3 (-3) & 5.3 (-3)\\
$^{61}$Fe & 33.0 & 4.24 & 4.2 (-3) & 1.7 (-3) &  & 6.4 (-2)\\
$^{61}$Co & 33.0 & 4.24 & 2.1 (-5) & 1.6 (-4) &  & 9.3 (-4)\\ 
$^{63}$Co & 33.0 & 4.24 & 3.8 (-2) & 1.6 (-2) &  & 1.4 (-2)\\
$^{59}$Mn & 220  & 5.39 & 1.1 (-2) & 2.2 (-2) & 7.2 (-1) & 1.4 (-1)\\ \hline
$^{58}$Co & 4.32 & 3.26 & 2.7 (-11) & 2.7 (-6) & 1.2 (-6) & 3.8 (-6)\\
$^{54}$Mn & 5.86 & 3.40 & 1.8 (-10) & 2.7 (-6) & 1.6 (-6) & 7.5 (-6)\\
$^{56}$Mn & 10.7 & 3.65 & 5.7 (-6) & 3.4 (-3) & 3.0 (-3) & 9.1 (-3)\\ 
$^{60}$Co & 10.7 & 3.65 & 8.3 (-7) & 6.6 (-4) & 1.4 (-3) & 3.4 (-3)\\
$^{50}$Sc & 33.0 & 4.24 & 6.6 (-4) & 1.2 (-2) & 2.8 (-2) & 1.8 (-1)\\ 
\end{tabular}
\end{table}


\begin{references}
\bibitem{Kla83}H.~V. Klapdor, Prog. Part. Nucl. Phys.~{\bf 10}, 131 (1983); Prog. Part. Nucl. Phys.~{\bf 17}, 419 (1986); AIP Conf. Proc.~{\bf 238}, 870 (1991).
\bibitem{Gro90}
K.~Grotz, H.V. Klapdor,
\newblock {\em The Weak Interaction in Nuclear, Particle and Astrophysics},
\newblock Adam Hilger, Bristol, Philadelphia, New York, 1990.
\bibitem{Sta90a}
A.~Staudt, E.~Bender, K.~Muto, H.~V. Klapdor-Kleingrothaus, Atomic Data and Nuclear Data Tables~{\bf 44}, 79 (1990).
\bibitem{Hir93}
M.~Hirsch, A.~Staudt, K.~Muto, H. V.~Klapdor-Kleingrothaus, Atomic Data and Nuclear Data Tables~{\bf53}, 165 (1993).
\bibitem{Ful80}
G.~M. Fuller, W.~A. Fowler, M.~J. Newman, Ap. J. Suppl.~{\bf 42}, 447 (1980);~{\bf 48},279 (1982); Ap. J.~{\bf 252}, 715 (1982).
\bibitem{Auf94}
M.~B. Aufderheide, I.~Fushiki, S.~E. Woosley, D.~H. Hartmann,
Ap. J. Suppl.~{\bf 91}, 389 (1994).
\bibitem{Auf96}
M.~B. Aufderheide, S.~D. Bloom, G.~J. Mathews, D.~A. Resler,
Phys. Rev. C.~{\bf 53}, 3139 (1996).
\bibitem{Oda94}
T.~Oda, M.~Hino, K.~Muto, M.~Takahara, K.~Sato,
Atomic Data and Nuclear Data Tables~{\bf 56}, 231 (1994).
\bibitem{Hal67}
J.~A. Halbleib, R.~A. Sorensen,
Nucl. Phys. A.~{\bf 98}, 542 (1967).
\bibitem{Kru84}
J.~Krumlinde, P.~M{\"o}ller,
Nucl. Phys. A.~{\bf 417}, 419 (1984).
\bibitem{Mut89}
K.~Muto, E.~Bender, H.V. {Klapdor--Kleingrothaus},
Z. Phys. A.~{\bf 334}, 187 (1989).
\bibitem{Sta89}
A.~Staudt, E.~Bender, K.~Muto, H.~V. Klapdor, Z. Phys. A~{\bf 334}, 47 (1989).
\bibitem{Kla84}
H.~V. Klapdor, J.~Metzinger, T.~Oda, Atomic Data and Nuclear Data Tables~{\bf 31}, 81 (1984).
\bibitem{Ben88}
E.~Bender, K.~Muto, H.~V. Klapdor,
Phys. Lett. B.~{\bf 208}, 53 (1988).
\bibitem{Hom96}
H.~Homma, E.~Bender, M.~Hirsch, K.~Muto, H.~V. Klapdor-Kleingrothaus, T.~Oda,
Phys. Rev. C.~{\bf 54}, 2972 (1996).
\bibitem{Mut92}
K.~Muto, E.~Bender, T.~Oda, H.~V. Klapdor,
Z. Phys. A.~{\bf 341}, 407 (1992).
\bibitem{Nab99a}
J.-U. Nabi, H.~V. Klapdor-Kleingrothaus,
Atomic Data and Nuclear Data Tables {\em {to appear in}}~{\bf 71}, (1999).
\bibitem{Lan98}
K.~Langanke, G.~Mart\'{\i}nez-Pinedo,
Phys. Lett. B.~{\bf 436}, 19 (1998); Preprint~{\bf nucl--th/9809082} (1998).
\bibitem{Mar98} G.~Mart\'{\i}nez-Pinedo, K.~Langanke, D.~J.~Dean, Preprint~{\bf nucl--th/9811095} (1998).
\bibitem{Nab99b}
J.-U. Nabi, H.~V. Klapdor-Kleingrothaus,
{\em in preparation}.
\bibitem{Nab99t}
J.-U. Nabi, Ph.D thesis, Heidelberg University, (1999).
\bibitem{Nab99c}
J.-U. Nabi, H.~V. Klapdor-Kleingrothaus, Proc. Second Internat. Conf. on Dark Matter in Astro and Particle Physics 1998, eds. H.V. Klapdor-Kleingrothaus and L. Baudis (IOP, Bristol and Philadelphia, 1999) (in press).
\bibitem{Nab99d}
J.-U. Nabi, H.~V. Klapdor-Kleingrothaus, Proc. Internat. Conf. on Nuclear Physics 1998, eds. R.Broda, B.Fornal, W.Meczynski (Acta Physica Polonica B, 1999) (in press).

\end{references}
\end{document}